\documentclass[10pt,a4paper]{article}
\usepackage[utf8]{inputenc}
\usepackage{amsmath}
\usepackage{amsfonts}
\usepackage{amssymb}
\usepackage{graphics}
\usepackage{graphicx}
\usepackage{epsfig}
\usepackage{subcaption}
\usepackage{amsmath,amssymb,graphicx} 

\begin{document}
\author{Sagardeep Talukdar, Riki Dutta,  Gautam Kumar Saharia  and\\ Sudipta Nandy \textsuperscript{a}\thanks{email: sudiptanandy@cottonuniversity.ac.in}  \\
 \\
 Cotton University, Guwahati, India}
	
\title{Bilinearization of the Fokas-Lenells equation Conservation laws and soliton interactions}

\maketitle	

\begin{center}
Abstract
\end{center}

In this paper, we propose the bilinearization of the Fokas-Lenells equation (FLE)  with a vanishing boundary condition. In the proposed bilinearization we make use of an  auxiliary function to convert the trilinear equations into a set of  bilinear equations. We  obtain bright $1$-soliton, $2$- soliton solutions and present the scheme for obtaining $N$ soliton solution. In the soliton solution the presence of an additional parameter allows tuning  the position of  soliton. We find that the proposed scheme of bilinearization using auxiliary function, considerably simplifies the procedure yet generates a more general solution than the one reported earlier. We  show that the obtained soliton solution reduces to an algebraic soliton in the limit of infinite width. Further we show explicitly  that the soliton interactions are elastic through asymptotic analysis, that is the amplitude of each soliton remains same before and after interaction. The mark of interaction  is left behind only in the phase of each soliton. Secondly, we propose a generalised Lax pair for the FLE and obtain the conserved quantities by solving Riccati equation. We believe that the present investigation would be useful to study the applications of FLE in nonlinear optics and other branches of physics.

%{In the end we found an important relation between the FL equation and the Landau-Lifshitz spin system through a suitable gauge transformation.}

\section{Introduction}
Fokas-Lenells equation  (FLE) \cite{Fokas, Lenells} is one of the four  categories of integrable equations which describes the propagation of ultrashort pulses in nonlinear medium. The other three are namely, nonlinear Schr\"odinger equation (NLSE) 
\cite{Hasegawa, Serkin00}, derivative NLSE (DNLSE) \cite{Kaup}, higher order NLSE (HNLSE) \cite{Hirota, SSE}. All four equations play significant role in the study of localised waves in nonlinear optical medium \cite{Agrawal, 
Kivshar, Serkin00,Serkin}. A common property in all these physical systems is the appearance of solitons, which arise as a result of a balance between the nonlinear and dispersive terms of the wave equations \cite{Agrawal}. In comparison to NLSE, the significance of FLE is due to the presence  of a spatio-temporal dispersion term in addition to the group velocity dispersion term. 
While the study on three other equations is done extensively, the study on  FLE is relatively  sparse.  
A few  notable contributions on FLE are
interaction between different localised waves
\cite{Ahmed}, solitary wave and elliptical solutions of FLE in presence of perturbation and modulation  instability \cite{Arshad}, combined optical solitary waves of FLE \cite{Triki2017},
dynamical behaviour of soliton solutions of FLE \cite{Hendi2021}, inverse transform of FLE with nonzero boundary condition \cite{Zhao}, optical soliton perturbation with FLE by different methods \cite{Krishnan2019, Aljohani2018}, derivation of dimensionless FLE with perturbation term \cite{Cinae2022}.  \\

% This criterion of inte-grability is extendable also to the quantum case. The Lax pair associated with the model is usualy a sign of such integrability,

\vspace{2mm}
The dimensionless form of FLE \cite{Fokas, Lenells} :
\begin{align} 
\label{FWDU}
iU_t  + a_1 U_{xx} + b|U|^2 U + a_2 (U_{xt}     
+ i b   |U|^2U_x) = 0 
\end{align}
where $U$ is  the  envelope function of an optical field. It can describe the ultrashort pulse propagation of an  optical field in a medium where the beam is allowed to diffract along one of the two, namely longitudinal and transverse directions.  An important fact about the FLE is that a gauge transformation of FLE belongs to the  hierarchy   of integrable DNLSE  \cite{Matsuno1, Matsuno2}.   Localised  solutions of FLE are obtained in different forms, namely  solitary waves \cite{Lenells},  'W'-shaped and other solitons \cite{Ghafri21}, soliton  solution  using inverse scattering transform \cite{Zhao}, using Darboux transformation \cite{Li2021},  using some other different methods \cite{Biswas2018}. N-soliton solution in   $\tau$- function is obtained in  \cite{Matsuno1}. 

In this manuscript our objective is to  propose an alternate simplified and systematic scheme, namely bilinearization by introducing an auxiliary function and obtain a generalized expression for multi-soliton solutions. 

%The soliton solution thus obtained has  been reduced to the earlier reported results. We analyse the properties of   $1$-soliton solution, $2$-soliton interactions. \\

Secondly, a limited class of soliton bearing equations exhibits an interesting property and belongs to the exclusive club of integrable systems \cite{Asokdas}. The most prominent definition of integrability is the integrability in the Liouville sense, that is  the existence of a set of infinite functionally independent conserved quantities, which are in involution \cite{Asokdas}. Importantly, the Poisson brackets of these conserved quantities with one another vanishes. 

Extraction of the  conserved quantities for the FLE and establishing the integrability of the  hierarchy in the Liouville sense remained unexplored till date. Thus  our second  objective is to obtain the same by solving  the Riccati equation for the $2\times 2$ dimensional Lax operator and consequently to obtain the whole hierarchy of conserved charges in a systematic way.

The structure of the manuscript is the following. In the following section we shall consider the  gauge transformed FLE.    
The bilinearization of FLE and one soliton solutions including algebraic soliton are described  in this section. Two Soliton interaction using asymptotic analysis  will be discussed in the third section.
Fourth section will cover the conserved quantities which  are obtained by using Lax pair and solving the Riccati equation.  Fifth section will be the concluding one.

\vspace{3mm}
\noindent
\section{Bilinearization of FLE with vanishing background} 
The  gauge transformation, 
\begin{align}
\label{GT}
U = \sqrt{\frac{m}{|b|}}n e^{i(n x +2 mn t)} u 
\end{align}
followed by transformation of variables,
\begin{align}
\label{VT}
\xi=x+m t; \quad \tau=-mn^2 t, 
\end{align}
where $n=\frac{1}{a_2} $ and $m=-\frac{a_1}{a_2}$.
converts eq. \ref{FWDU} into

\begin{align} 
\label{FLE}
u_{\xi \tau} = u - 2i \sigma |u|^2 u_\xi      
\end{align}
    
Under  the vanishing background condition $|u| \rightarrow 0$ as $ |\xi| \rightarrow \pm \infty $ we expect a bright soliton solution by applying  the bilinearization method.
 
%The simplest form of solution of  eq. (\ref{FLE}) is the plane wave. A plane wave solution, obtained as ansatz is 
%\begin{align}\label{PW}
%q = A e^{k \ x + \frac{1}{k} \ t + |A|^2} 
%\end{align} 
%where $k$ is the wave vector and $\omega=\frac{1}{k}$ is the frequency and $A$ is the amplitude of the plane wave. 

%A localised solution in the form of bright soliton or dark soliton is  obtained  by applying a direct method, namely the Hirota bilinear method.

To write  eq. \ref{FLE} in the bilinear form let us assume 

\begin{align}
\label{bilin0}
q =\frac{g}{f}
\end{align}
where $g$ and $f$ are two complex functions of ($\xi,\ \tau$).  Using eq. \ref{bilin0}  and  eq. \ref{FLE} we obtain a nonlinear  equation in terms of newly introduced fields,
\begin{align}
\label{Bilin}
\frac{1}{f^2}(D_\xi D_\tau -1) g.f - \frac{g}{f^3}D_\xi D_\tau(f.f) +  \frac{2i|g|^2}{f^3 f^*}D_\xi (g.f) + \frac{s |g|^2}{f^3} - \frac{s |g|^2 f^*}{f^3 f^*} =0
\end{align} 

where $D_\xi$, $D_\tau $ are the  Hirota derivatives\cite{Hirota, Nandy2015PRE}, which   are defined as

\begin{align}
D_\xi^m D_\tau^n g(\xi,\tau).f(\xi,\tau)= 
(\frac{\partial}{\partial \xi}  - \frac{\partial}{\partial \xi^\prime})^m
(\frac{\partial}{\partial \tau}  - \frac{\partial}{\partial \tau^\prime})^n
g(\xi,\tau).f(\xi^\prime,\tau^\prime)\Bigg|_{ (\xi=\xi^\prime)(\tau= \tau^\prime)}
\end{align} 

Notice that the last two terms in eq. \ref{Bilin} contain  an  auxiliary function $s$, which is introduced  so that  eq. \ref{Bilin} can be cast into following bilinear equations in terms of $g$, $f$ and $s$.
\begin{align}
\label{BR1}
(D_\xi D_\tau -1)g.f&=0\\
\label{BR2}
D_\xi D_\tau (f.f)  &= sg^*\\
\label{BR3}
2i D_\xi (g.f)   &= sf^* 
\end{align}
To obtain the soliton solution,  $g$ and $f$  are expanded with respect to an arbitrary
parameter $\epsilon$ as follows: 

\begin{align}
\label{GF}
g&= \epsilon g_1 + \epsilon^3 g_3 +..... \\
f&= f_0 + \epsilon^2 f_2 + \epsilon^4 f_4 +.....
\end{align} 
The auxiliary function is expanded as  

\begin{align}
\label{S}
s= \epsilon s_1 + \epsilon^3 s_3 + .....
\end{align}  
    
\subsection{Bright soliton}
The one soliton solution (1-SS) of eq. \ref{FLE} is obtained  by using eqs. \ref{bilin0}, \ref{GF} -  \ref{S}, subsequently  
dropping terms of order greater than or equal to  $\epsilon^3 $ in  $g$ and $f$ and $s$. That is, 

\begin{align}
\label{sol1}
q= \frac{\epsilon g_1}{f_0 + \epsilon^2 f_2}\Big|_{\epsilon=1} 
\end{align}

Let us consider 
\begin{align}
\label{G1}
g_1&= \alpha  e^{\theta (\xi, \tau)   } \\
\label{F1}
f_0 &=\beta; \quad f_2 = \beta_2 e^{\theta (\xi, \tau) +  \theta^* (\xi, \tau)}\\
\label{S1}
s&= c e^{\theta(\xi, \tau)}
\end{align} 
substituting eqs. \ref{sol1} -\ref{S1} in eqs. \ref{BR1}- \ref{BR3} we obtain the 1-SS,  where the parameters are 

\begin{align}
\theta &= p \ \xi + \frac{1}{p} \tau + \theta_0  \\ 
\beta_2 &= i \frac{|\alpha|^2 |p|^2 p}{\beta^* (p+p^*)^2 }; 
\quad \quad c = \frac{2i\beta\alpha p}{\beta^*} 
\end{align}
$\alpha$, $\beta$ and $p$ are arbitrary complex constants. $\theta_0$ is also a complex constant representing the initial phase.
 
\begin{figure}
	\centering
	\includegraphics[width=0.5\linewidth]{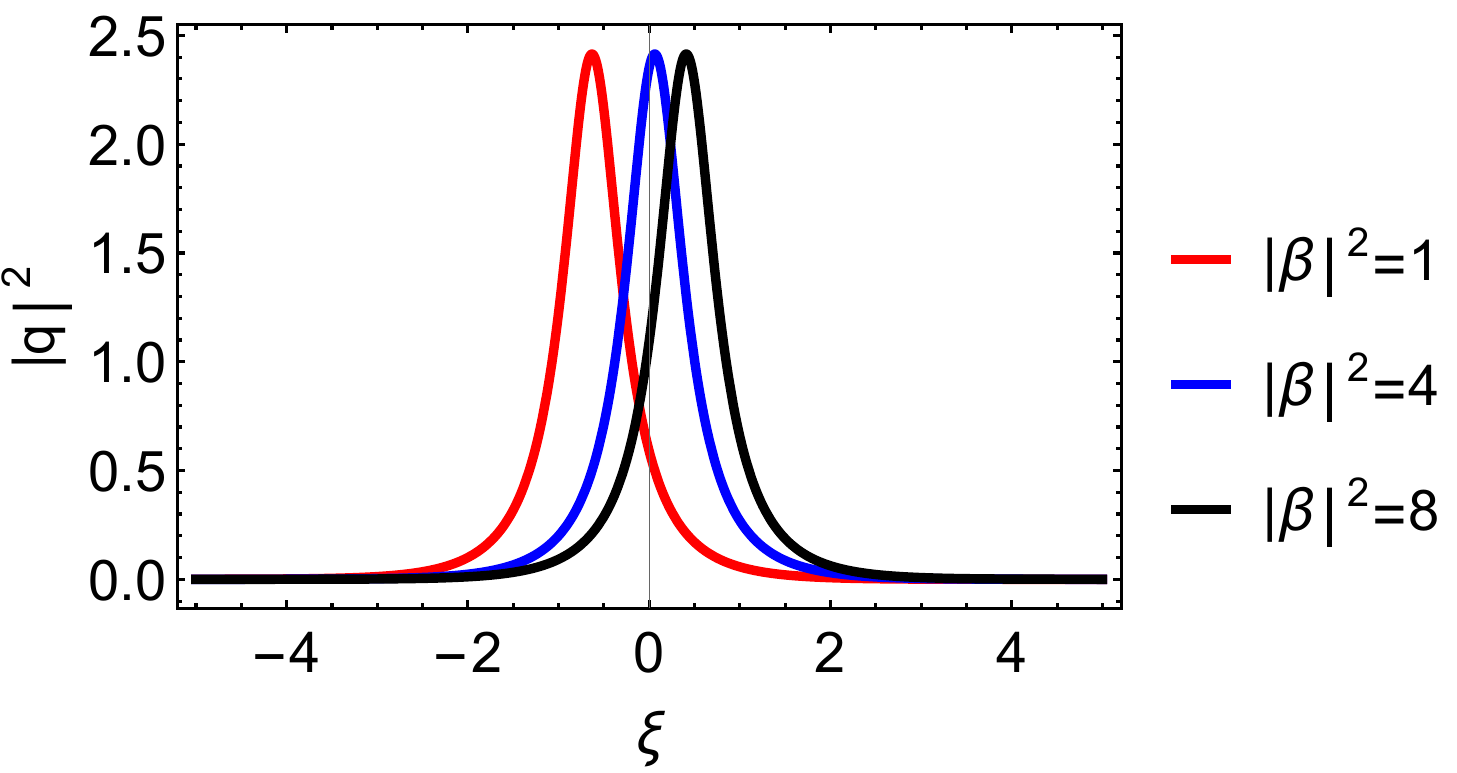}
	\caption{Shift in the soliton position with respect to $\beta$.}
	\label{fig:beta1}
\end{figure}

Notice that due to the introduction of an auxiliary function  the  bilinearization process becomes considerably straightforward  as compared to the earlier reported methods \cite{Zhao, Matsuno1}. The additional parameter $\beta$ introduced in the solution is  responsible for the shift in the central position of the soliton. Other than that $\beta$ does not interfere with other soliton properties. Figure \ref{fig:beta1} shows the shift in the position of the soliton with change in $\beta$.  Notice that for $\alpha = -(p+p^*) $ and $\beta= -i |p|^2$ eq.  \ref{sol1}   reduces to the soliton solution obtained in \cite{Matsuno1}.  

\begin{figure*}[htbp]
	\begin{subfigure}{.5 \textwidth}
		\centering
		\includegraphics[width=.9 \textwidth]{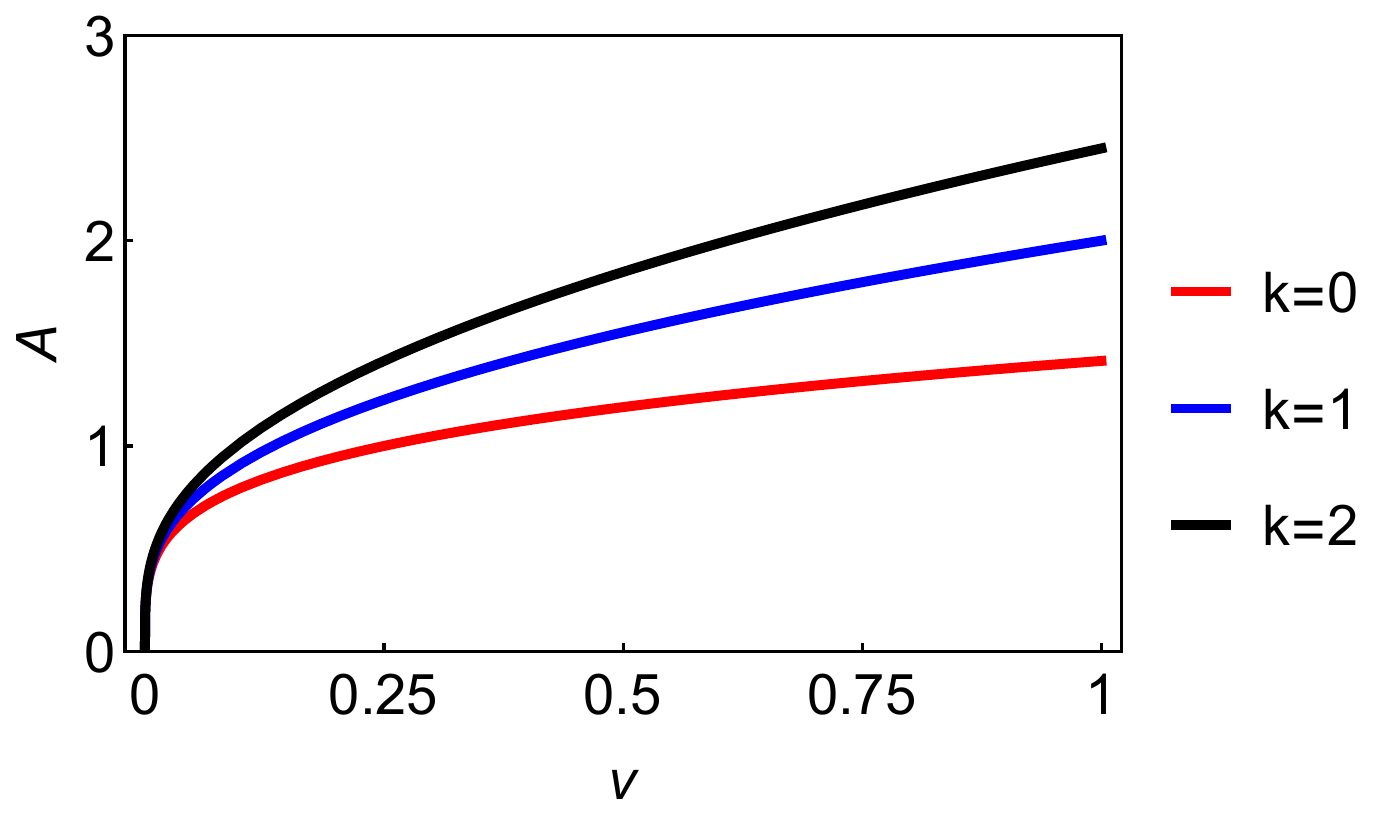}
		\caption{}
		\label{fig:amp-vel-1}
	\end{subfigure}%
	\begin{subfigure}{.5 \textwidth}
		\centering
		\includegraphics[width=.9\textwidth]{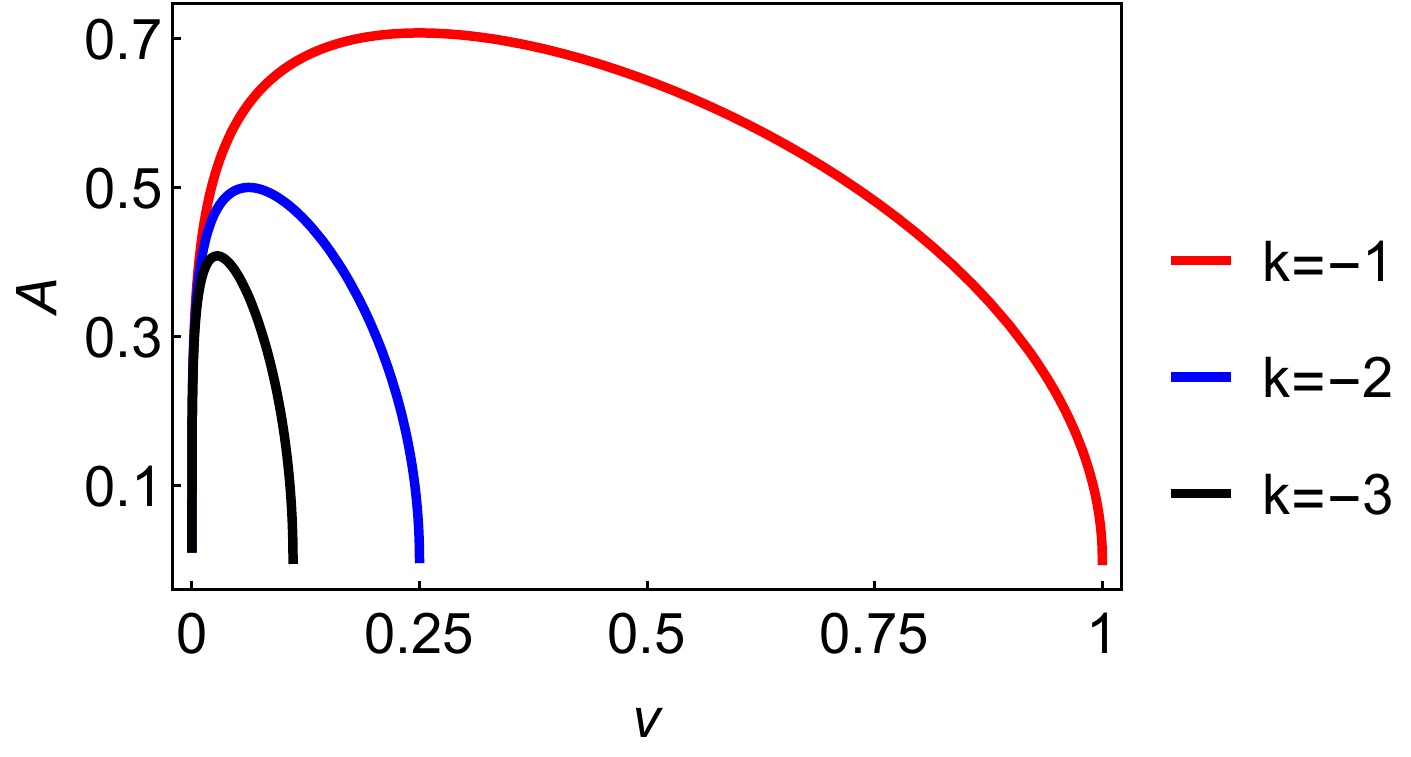}
		\caption{}
		\label{fig:amp-vel-2}
	\end{subfigure}%
	\caption{(a) Variation of amplitude with velocity for positive values of wave vector (k) \ (b) Variation of amplitude with velocity for negative values of wave vector }
	\label{fig:amp-vel}
\end{figure*}

We may further identify  the following physical quantities associated with the soliton wave, namely wave number, 
$k = \frac{-i}{2}(p-p^*)$,   the    frequency shift,  
$ \Omega(k) = \frac{\frac{i}{2}(p-p^*)}{|p|^2 }$,  
the soliton velocity, $ v= -\frac{d\Omega}{dk}= \frac{1}{|p|^2}$ and the width inverse, $a=\frac{1}{2}(p+p^*)$. Further we may write the amplitude  in terms of velocity as $ A = \sqrt{v}    \sqrt{ 2(\frac{1}{\sqrt{v}} +k)}$.   
Notice that the amplitude  vs  velocity is not linear as in the case of conventional NLSE soliton \cite{Hasegawa} but has an interesting relationship where the amplitude of soliton  is a function of  two of the  following  parameters, '$k$, $v$ and $a$'. The same is  illustrated in figure \ref{fig:amp-vel} (a,\ b). \\  

Figure \ref{fig:amp-vel}  shows amplitude ($A$) \textit{vs}  velocity ($v$) graph for (a) $ k = 0,1,2$ and (b) $ k = -1,-2,-3$. It shows that for positive values of $k$, the amplitude increases monotonically with the increase in $v$. For negative values of $k$, the relation is not monotonic. The amplitude increases initially and reaches a maximum value at $ v= \frac{1}{4b^2} $ and then decreases.   

\begin{figure*}[htbp]
	\begin{subfigure}{.5 \textwidth}
		\centering
		\includegraphics[width=.9 \textwidth]{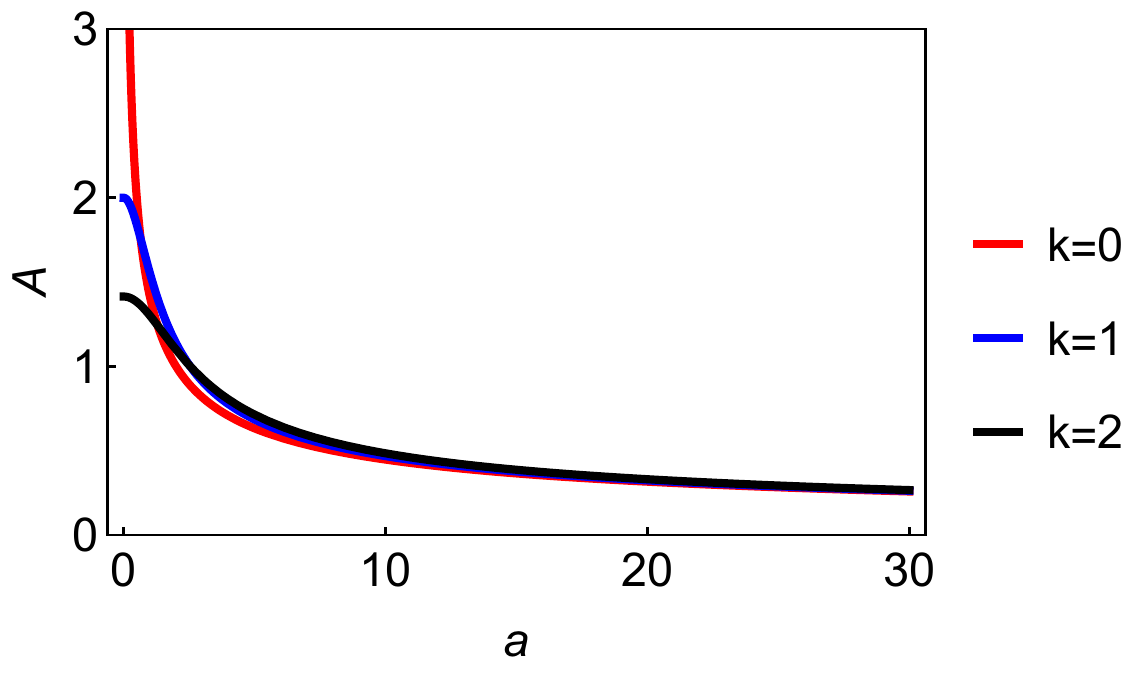}
		\caption{}
		\label{fig:amp-width-1}
	\end{subfigure}%
	\begin{subfigure}{.5 \textwidth}
		\centering
		\includegraphics[width=.9\textwidth]{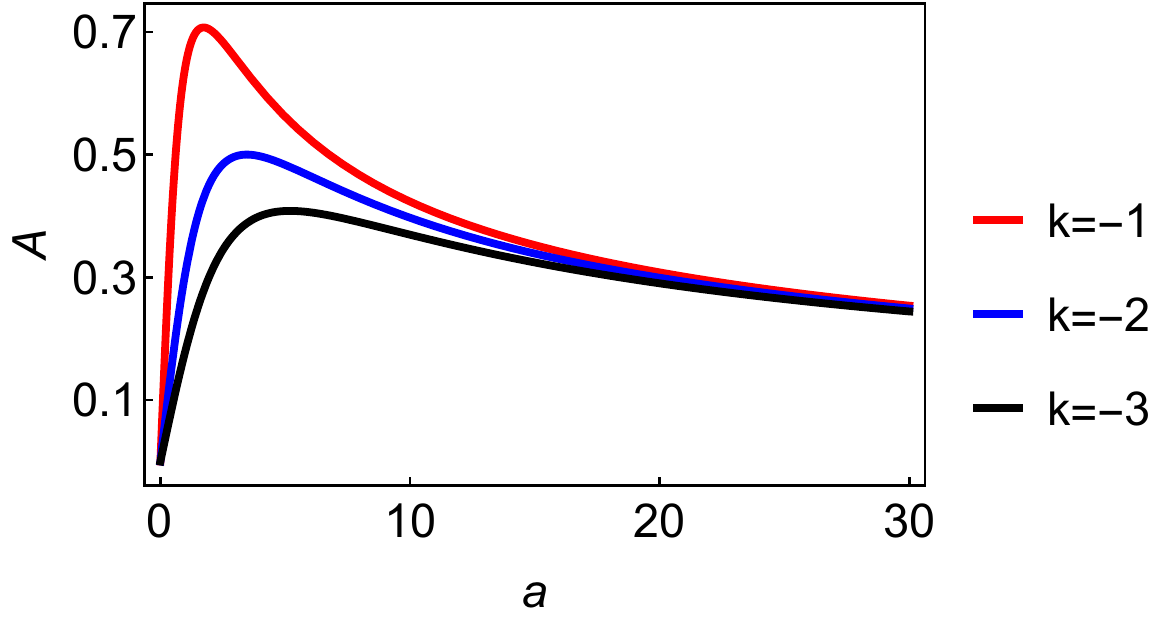}
		\caption{}
		\label{fig:amp-width-2}
	\end{subfigure}%
	\caption{(a) Variation of amplitude with width inverse for positive values of wave vector (k) \ (b) Variation of amplitude with width inverse for negative values of wave vector }
	\label{fig:amp-width}
\end{figure*}

Figure \ref{fig:amp-width} shows amplitude (A) \textit{vs} width inverse ($a$) graphs for (a) $ k = 0,1,2$ and (b) $ k = -1,-2,-3$. Notice that for positive values of $k$, the amplitude decreases monotonically with the increase in $a$. For negative values of $k$, the relation is however not monotonic. The amplitude increases initially and reaches a maximum value at $a= -\sqrt{3}b$ and then decreases.
 
The above mentioned properties are quite unique and in contrast to that of the conventional NLSE soliton. 

\subsection{Algebraic Soliton}

\begin{figure*}[htbp]
	\begin{subfigure}{.5 \textwidth}
		\centering
		\includegraphics[width=.9 \textwidth, height=0.2\textheight]{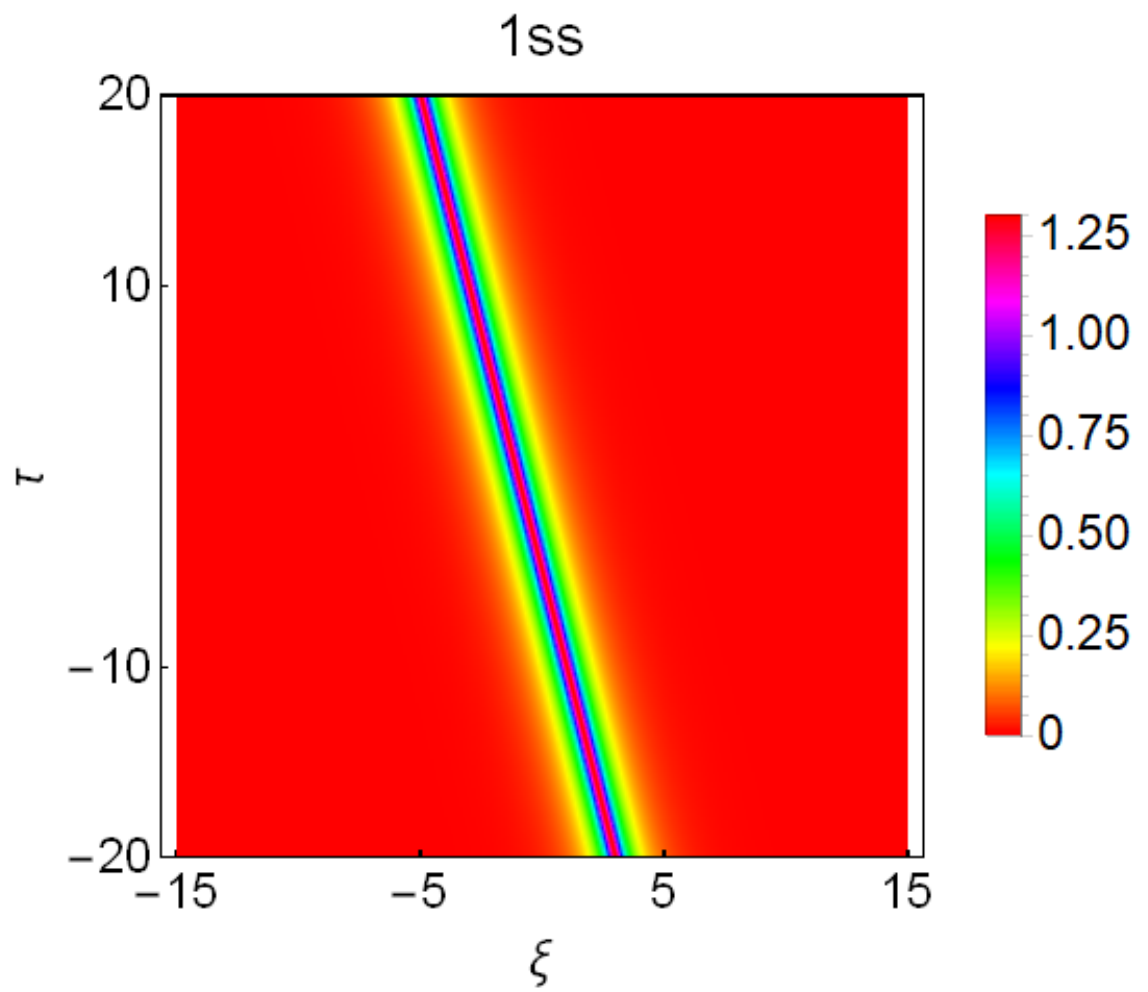}
		\caption{}
		\label{fig:1ss}
	\end{subfigure}%
	\begin{subfigure}{.5 \textwidth}
		\centering
		\includegraphics[width=1\textwidth, height=0.19\textheight]{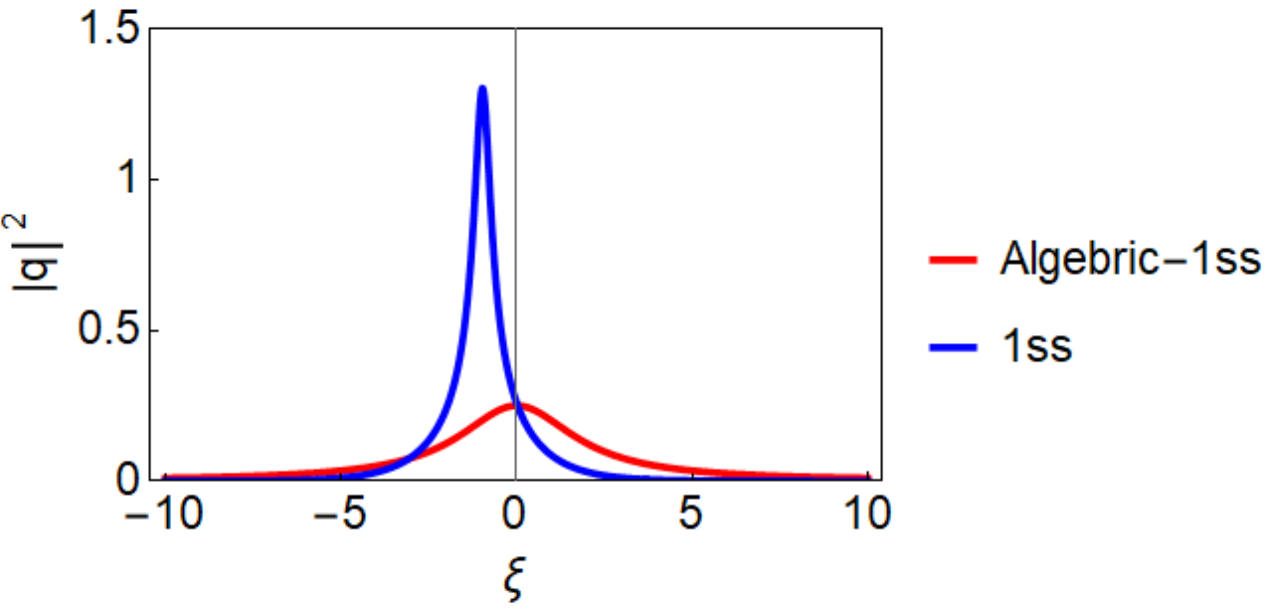}
		\caption{}
		\label{fig:Algebric1ss}
	\end{subfigure}%
	\caption{ Density plot of (a) one soliton in component $|q|$ with $a=1$, $k=2$, $\alpha=1+2 i$ and $\beta=1+i$. \ (b) 2D plot of a normal soliton given in (a) and an algebraic soliton with $ \ k =2 $. }
	\label{fig-FL1}
\end{figure*}

One  unique feature of FLE soliton is that in the limit of infinite width ($a\rightarrow 0$) FLE soliton does not vanish but the amplitude reduces to a finite value. To show this let us consider $ \beta= {p^*}^2$,   
$\alpha = (p+p^*) $ and the phase $\theta_{0} = a x_0 + 2\delta$ where $e^{2\delta}=  k $. The 1-SS from eq. \ref{sol1} then reduces to the form,   
	
\begin{align}
q=\frac{- e^{i\chi(\xi,\tau)}}{( \xi + v \tau + x_0  )   - i k}      \label{Algebric}   
\end{align} 
where 
$\chi(\xi,\tau) = k\  \xi-\frac{\tau}{k} + \chi_0 $,  \ \
$\chi_0$ is a constant phase of the oscillating function.\\
Eq. \ref{Algebric} is nothing but the expression of an algebraic soliton,  where the envelope function disappears giving rise to  an algebraic form.  However  the soliton properties are still  maintained. In the limit $a\to 0$, \  the amplitude of the soliton reduces to a non-zero finite value  and is proportional to $\sqrt{v}$. This is unlike conventional  NLSE soliton, where the amplitude tends to zero in the limit of infinite width ($a\rightarrow0$). Further notice that in the process of algebraic reduction two of the  free parameters, $\alpha $ and $\beta$ have been constrained and thus leaving behind only one free parameter, namely the wave vector $k$. Figure \ref{fig-FL1}   shows (a) the density plot of a soliton with $a=1$,\ $k =2$,\  $ \alpha= 1+2i$,   \ $ \beta=1+i $ (b) 2D plot of   an algebraic soliton with $k =2$, in comparison to the soliton (a).

\section{Two Soliton}   

The two soliton solution ($2$-SS) of FLE eq. \ref{FLE}  is obtained from eq. \ref{bilin0} by dropping terms of  order greater than or equal to  $\epsilon^5 $ in  $g$, $f$ and $s$ in eqs. \ref{GF}-\ref{S}.

\begin{align}
\label{2sol}
q= \frac{\epsilon g_1 + \epsilon^3 g_3}{ f_0 + \epsilon^2 f_2 + \epsilon^4 f_4}\Big|_{\epsilon=1} 
\end{align} 
 
Let us consider, 

\begin{align}
\label{G2}
 g_1 &= \alpha_1 e^{\theta_1(\xi,\tau) }+ \alpha_2 e^{\theta_2(\xi,\tau) }\\
g_3 &= \alpha_3 e^{\theta_1(\xi,\tau)+\theta^*_1(\xi,\tau)+\theta_2(\xi,\tau) } 
+\alpha_4 e^{\theta_1(\xi,\tau)+\theta_2(\xi,\tau)+\theta^*_2(\xi,\tau) } \\
f_0&= \beta; \quad \ f_2= \beta_2 e^{\theta_1(\xi,\tau)+\theta^*_1(\xi,\tau)}  +  
\beta_3 e^{\theta_2(\xi,\tau)+\theta^*_2(\xi,\tau)} \nonumber \\ &+ 
\beta_4 e^{\theta_1(\xi,\tau)+\theta^*_2(\xi,\tau)} +
\beta_5 e^{\theta^*_1(\xi,\tau)+\theta_2(\xi,\tau)}\\
f_4 & =\beta_6
e^{\theta_1(\xi,\tau)+\theta^*_1(\xi,\tau)+\theta_2(\xi,\tau)+\theta^*_2(\xi,\tau)} 
\end{align}
\begin{align}
\label{S2}
s&= c_1 e^{\theta_1(\xi,\tau)}+ c_2 e^{\theta_2(\xi,\tau)} + c_3 e^{\theta_1(\xi,\tau)+\theta^*_1(\xi,\tau)+\theta_2(\xi,\tau)}\nonumber \\ 
& + c_4 e^{ \theta_1(\xi,\tau)+\theta_2(\xi,\tau)+\theta^*_2(\xi,\tau)}
\end{align}

Substituting eqs. \ref{2sol}- \ref{S2} in eqs. \ref{BR1} - \ref{BR3}
we obtain the following  parameters,
\begin{align*}
\theta_1 &= p_1 \ \xi + \frac{1}{p_1} \tau + \theta_{10};
\quad \quad
\theta_2 = p_2 \ \xi + \frac{1}{p_2} \tau + \theta_{20}  \\
\alpha_3 &=\frac{i |\alpha_1|^2 \alpha_2 {p^*_1}^3 (p_1-p_2)^2}{|\beta|^2 (p_1+p^*_1)^2 (p^*_1+p_2)^2};
\alpha_4 =\frac{i |\alpha_2|^2 \alpha_1 {p^*_2}^3 (p_1-p_2)^2}{|\beta|^2 (p_2+p^*_2)^2 (p_1+p^*_2)^2}\\
\beta_2&=\frac{i |\alpha_1|^2 |p_1|^2 p_1}{\beta^* (p_1 + p^*_1)^2};
\quad\quad\quad\hspace{0.5em}
\beta_3=\frac{i |\alpha_2|^2 |p_2|^2 p_2}{\beta^* (p^*_2 + p_2)^2}\\
\beta_4&=\frac{i \alpha^*_2 \alpha_1 {p_1}^2 p^*_2}{\beta^* (p_1 + p^*_2)^2};
\quad\quad\quad\hspace{0.5em}
\beta_5=\frac{i \alpha_2 \alpha^*_1 p_2^2 p_1^*}{\beta^* (p_2 + p^*_1)^2}\\
\beta_6&=\frac{- p_1 p_2 |p_1|^2 |p_2|^2 |\alpha_1|^2 |\alpha_2|^2  |(p_1-p_2)|^4}{\beta^* |\beta|^2 (p_1+p^*_1)^2 (p^*_1+p_2)^2 (p_1+p^*_2)^2 (p_2+p^*_2)^2}
\end{align*}      
$\beta$, $\alpha_1$ ,$\alpha_2$ and $p_1$ ,$p_2$ are arbitrary complex constants.

To study the amplitudes and the phase shift of $2$-soliton we use asymptotic analysis.

\vspace{3mm}

\subsection{Asymptotic analysis} 

When asymptotically  apart from each other multi-solitons are essentially separated  single solitons  \cite{Nandy2015PRE}. For asymptotic analysis let us consider  two bright solitons $q_1$ and $q_2$ which interact as they move with a relative velocity.\\

Before interaction as $\tau \rightarrow -\infty $,

\begin{align}
&q_1= \frac{\alpha_1 e^{p_1 \xi+\frac{\tau}{p_1}}}{\beta + \beta_2 e^{(p_1+p^*_1)\xi+(\frac{1}{p_1}+\frac{1}{p^*_1})\tau}}; \\
& \theta_1(\xi,\ \tau) \to 0; \quad \theta_2(\xi,\ \tau) \to -\infty \nonumber \\
%&(p_1+p^*_1)x+(\frac{1}{p_1}+\frac{1}{p^*_1})t\to 0, (p_2+p^*_2)x+(\frac{1}{p_2}+\frac{1}{p^*_2})t \to -\infty \\
&q_2= \frac{\alpha_3 e^{p_2 \xi+\frac{\tau}{p_2}}}{\beta_2+ \beta_6 e^{(p_2+p^*_2)\xi+(\frac{1}{p_2}+\frac{1}{p^*_2})\tau}}; \\
&\theta_2(\xi,\ \tau) \to 0; \quad \theta_1(\xi,\ \tau) \to \infty \nonumber  
%&(p_2+p^*_2)x+(\frac{1}{p_2}+\frac{1}{p^*_2})t\to 0, (p_1+p^*_1)x+(\frac{1}{p_1}+\frac{1}{p^*_1})t \to \infty \\
\end{align}

Consequently,

\begin{align}
\label{Q1int}
&|q_1|^2=\frac{2 a^2_1}{(a^2_1+b^2_1)^{3/2}}\times \frac{1}{cosh[2(\theta_1+ln\ \delta_1)]-\frac{b_1}{\sqrt{a^2_1+b^2_1}}}\\
\label{Q2int}
&|q_2|^2=\frac{2 a^2_2}{(a^2_2+b^2_2)^{3/2}}\times \frac{1}{cosh[2(\theta_2+ln\ \rho_2)]-\frac{b_2}{\sqrt{a^2_2+b^2_2}}}
\end{align}

after interaction as $\tau \to \infty$,
\begin{align}
&q_1= \frac{\alpha_4 e^{p_1 \xi+\frac{\tau}{p_1}}}{\beta_3+ \beta_6 e^{(p_1+p^*_1)\xi+(\frac{1}{p_1}+\frac{1}{p^*_1})\tau}}; \\
& \theta_1(\xi,\ \tau) \to 0; \quad \theta_2(\xi,\ \tau) \to \infty \nonumber  \\
%&(p_1+p^*_1)x+(\frac{1}{p_1}+\frac{1}{p^*_1})t\to 0, (p_2+p^*_2)x+(\frac{1}{p_2}+\frac{1}{p^*_2})t \to \infty\\
&q_2= \frac{\alpha_2 e^{p_2 \xi+\frac{\tau}{p_2}}}{\beta+ \beta_3 e^{(p_2+p^*_2)\xi+(\frac{1}{p_2}+\frac{1}{p^*_2})\tau}}; \\
&\theta_2(\xi,\ \tau) \to 0; \quad \theta_1(\xi,\ \tau) \to -\infty \nonumber 
%&(p_2+p^*_2)x+(\frac{1}{p_2}+\frac{1}{p^*_2})t\to 0, (p_1+p^*_1)x+(\frac{1}{p_1}+\frac{1}{p^*_1})t \to -\infty \\ 
\end{align}

Consequently,

\begin{align}
\label{Q1Aint}
&|q_1|^2=\frac{2 a^2_1}{(a^2_1+b^2_1)^{3/2}}\times \frac{1}{cosh[2(\theta_1+ln\ \rho_1)]-\frac{b_1}{\sqrt{a^2_1+b^2_1}}}\\
\label{Q2Aint}
&|q_2|^2=\frac{2 a^2_2}{(a^2_2+b^2_2)^{3/2}}\times \frac{1}{cosh[2(\theta_2+ln\ \delta_2)]-\frac{b_2}{\sqrt{a^2_2+b^2_2}}}
\end{align}
where, $\theta_1=a_1 \xi+\frac{a_1 \tau}{a^2_1+b^2_1}$, $\theta_2=a_2 \xi+\frac{a_2 \tau}{a^2_2+b^2_2}$, $\delta_1=\frac{|\alpha_1| (a^2_1+b^2_1)^{3/4}}{2 a_1 |\beta|}$,$\delta_2=\frac{|\alpha_2| (a^2_2+b^2_2)^{3/4}}{2 a_2 |\beta|}$ \\
$\rho_1=\delta_1 \ \frac{(a_1+a_2)^2+(b_1-b_2)^2}{(a_1-a_2)^2+(b_1-b_2)^2}$ and $\rho_2=\delta_2 \ \frac{(a_1+a_2)^2+(b_1-b_2)^2}{(a_1-a_2)^2+(b_1-b_2)^2}$. \\
\\

\begin{figure}
	\centering
	\includegraphics[width=0.7\linewidth]{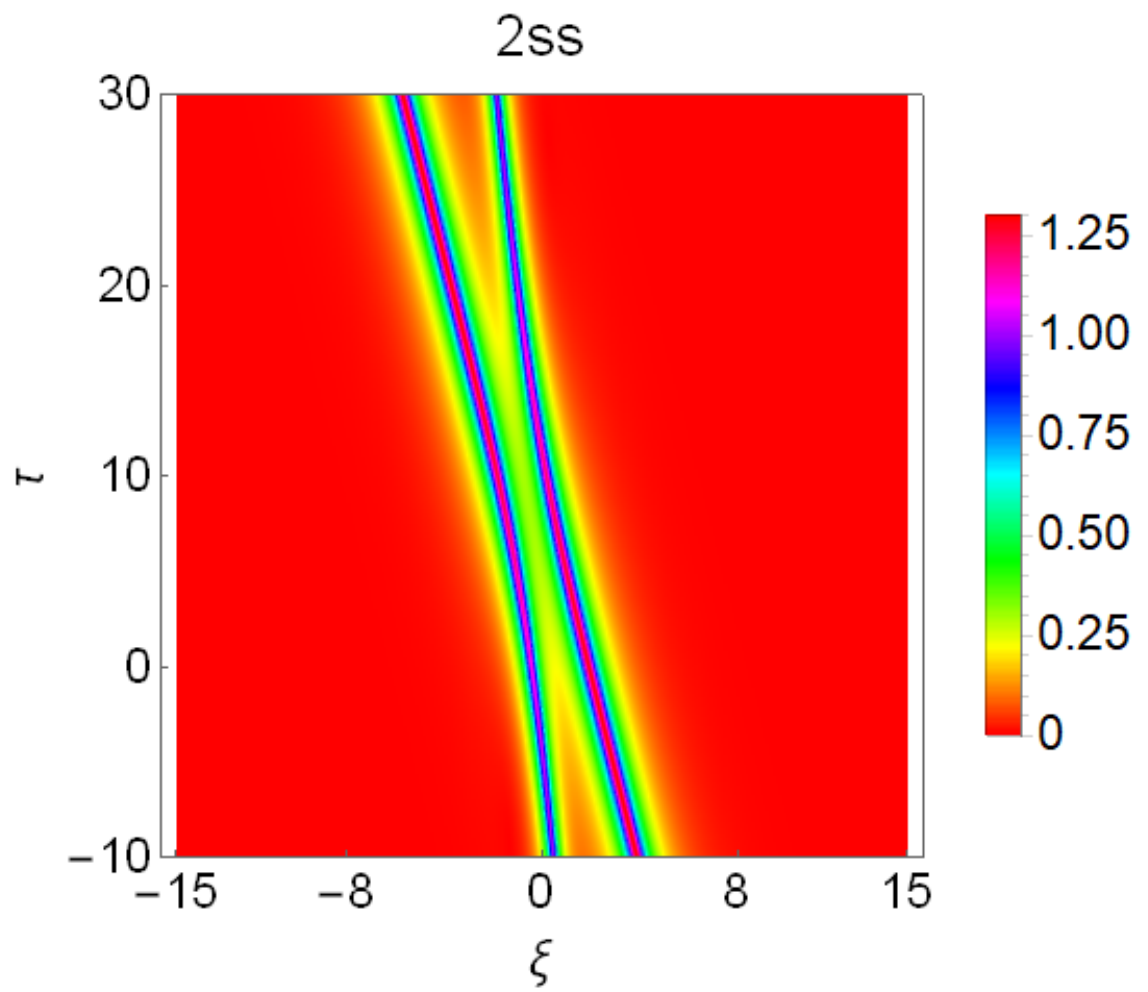}
	\caption{Interaction of 2 solitons moving with a relative velocity and having parameters $a_1=1$,\ $a_2=2$,\ $k_1=2$,$k_2=3$,\ $\alpha_1=1+i$,\ $\alpha_2=3+2i$ and $\beta=1+3i$.}
	\label{fig2ss-1}	
\end{figure}

Notice that the amplitude of soliton $q_1$ evaluated from eq. \ref{Q1int} and eq. \ref{Q1Aint} are found to be same. Similarly  the  amplitude of soliton $q_2$ evaluated from eq. \ref{Q2int} and eq. \ref{Q2Aint} are found to be same. That is the amplitude of solitons before and after interaction remain same, which is one of the important characteristics of soliton. However, the position of each soliton changes as a result of interaction, as evident from the  figure \ref{fig2ss-1}. The same can be calculated from
expressions in eqs. \ref{Q1int},\  \ref{Q1Aint}  and eqs. \ref{Q2int}, \ref{Q2Aint} and the magnitude of shift is given by 
\begin{align}
|\Delta_{ph}| =ln(\frac{(a_1-a_2)^2 + (k_1-k_2)^2}{(a_1+a_2)^2 + (k_1-k_2)^2})
\end{align}
Indeed, the phase shifts are opposite to each other. 

Proceeding in a similar way, $N$-soliton solution  is obtained by dropping terms of  order greater than or equal to  $ \epsilon^{2N+1}$ in eqs. \ref{GF} - \ref{S}, which is

\begin{align}
q= \frac{\epsilon g_1 + \epsilon^3 g_3 +.....+\epsilon^{2N-1} g_{2N-1}}{ f_0 + \epsilon^2 f_2 +.....+ \epsilon^{2N} f_{2N}}\Big|_{\epsilon=1} 
\end{align}
where $g_i$, $f_i$ and $s_i$ are expanded as prescribed 
in \cite{Nandy2021Chaos, Nandy2019}.

In the next section we will show the existence of infinite conserved quantities for the integrable FLE.

\section{Conserved Quantities} 

In order to establish integrability in the Liouville sense, that is to compute the infinite number of conserved quantities we first find the Riccati equation from the Lax equation. The linear transformation equation for FLE in terms of  Lax pair is given by

\begin{align}
\label{LaxFL}
\partial_\xi {\bf \Psi} = L {\bf \Psi} \\
\partial_\tau {\bf \Psi} = M {\bf \Psi}
\end{align}

where $\bf{\Psi}$ is a two component vector field, expressed as
\begin{align}
\label{Psi}
{\bf \Psi}= ({\Psi_1} \ {\Psi_2})^T
\end{align}

and $L$ and $M$ are given by
\begin{align}
\label{Lax2}
L = \frac{-i\zeta^a }{2} \ \Sigma   + \zeta^b \  \partial_\xi u           \\
M= \frac{i }{2  } \zeta^c \ \Sigma -  {i}{\zeta^d}\ \Sigma u + i\Sigma u^2
\end{align} 

where $\zeta $   is the spectral parameter. To satisfy the zero curvature condition, namely
\begin{align}
L_\tau -M_\xi + [L,M] = 0
\end{align} 
the exponents of 
$\zeta $, namely $a,\ b,\ c,\  d $ should satisfy the following relations,

\begin{align}
a = b-d; \quad 2d = c; \quad b =-d
\end{align} 

$\Sigma$  and $u$ are $2\times 2 $ matrices, defined as follows

$ \quad \Sigma = \left(  \begin{tabular}{c c}
%\hline 
1 & 0 \\ 
%\hline 
0 & -1  \\ 
%\hline 
\end{tabular} \right) , \quad \quad
 u = \left(  \begin{tabular}{c c}
%\hline 
0 & $q$ \\ 
%\hline 
$-q^*$ & 0   \\
%\hline 
\end{tabular} \right) $ \\

Writing the first  Lax eq. \ref{Lax2} in  component form.
\begin{align}
\label{component1}
\Psi_{1\xi} & = \frac{-i \zeta^2}{2} \Psi_1  + \zeta \ q_\xi \ \Psi_2\\
\label{component2}
\Psi_{2\xi} & = \frac{i \zeta^2}{2} \Psi_2 -  \zeta \ q_\xi^* \ \Psi_1  
\end{align}

Now following a similar procedure as in \cite{Ghosh1999} we write 
\begin{align}
\Gamma = \frac{\Psi_1}{\Psi_2}
\end{align}

Then from eqs. \ref{component1}, \ref{component2}  we obtain a first order  nonlinear differential equation,
\begin{align}
\label{Riccati2}
\Gamma_\xi = -i \zeta^2 \Gamma + \zeta q_\xi 
+ \zeta q_\xi^* \Gamma^2
\end{align}
which is known as Riccati equation. The solution of the Riccati eq.  \ref{Riccati2} is related to the conserved quantities   in the following way, 

\begin{align}
\label{Conserve-1}
 ln(a_{22}(\zeta))  =  ln (e^{- i \frac{\zeta^2}{2} \xi} \Psi_2 )|_{\xi \rightarrow \pm \infty} 
 = -\zeta \int_{-\infty}^{\infty} q_\xi^* \Gamma \ d\xi= \sum_{n=0}^{\pm \infty} i^{n} H_n \zeta^{-n+1}   
\end{align}

In eq. \ref{Conserve-1}, $ a_{22}$ is the scattering parameter and is time independent. In the summation, the negative powers of $\zeta $ give  positive hierarchy $H_n$,  where as the positive powers of $\zeta $ give negative hierarchy $ H_{-n}$.  \\
Consider the eq. \ref{Riccati2} has a series solution, 

\begin{align}
\Gamma = \sum_{n=0}^{\infty} a_{\pm n} \zeta^{\mp n}
\end{align}

Substituting $\Gamma$ in eq. \ref{Riccati2} we obtain the coefficients $ a_n $; 

\begin{align}
a_0 & =0;\quad  a_1= - i \partial_\xi q \\
a_2 &=0; \quad a_3 =q_{\xi \xi} + i |q_\xi|^2 q_\xi\\
a_4 &=0; \quad   a_5= i (q_{\xi \xi \xi}-2|q_\xi|^4 q_x ) 
-4|q_\xi|^2q_{\xi \xi} - (q_\xi)^2 q_{\xi \xi}^* \\
&\cdots \qquad \cdots \qquad \cdots \qquad \cdots \nonumber \\
a_{-1}&= q; \quad  a_{-2}=0 ;\quad  a_{-3}=\int_{-\infty}^{\infty}( -iq+q^2 q^*_\xi )d\xi\\
&\cdots \qquad \cdots \qquad \cdots \qquad \cdots \nonumber
\end{align}

The conserved quantities thus obtained are

\begin{align}
H_1 &= \int_{-\infty}^{\infty} |q_\xi|^2 d\xi \\
H_3 &= \int_{-\infty}^{\infty} ( |q_\xi|^4   -i q_\xi^* q_{\xi \xi} ) 
  d\xi  \\
H_5 &= \int_{-\infty}^{\infty} (- q_\xi^* q_{\xi\xi\xi} + 2|q_\xi|^6 -3i |q_\xi|^2 q_\xi^* q_{\xi\xi}  ) d\xi \\
&\cdots \qquad \cdots \qquad \cdots \qquad \cdots \nonumber \\
&\cdots \qquad \cdots \qquad \cdots \qquad \cdots \nonumber 
\end{align}

\begin{align}
\label{negative-1}
H_{-1} &= \int_{-\infty}^{\infty}- i q_\xi^*q\  d\xi \\
\label{negative-2}
H_{-3} &= \int_{-\infty}^{\infty} -( |q|^2  +
i |q|^2q q_\xi^*)\  d\xi  \\
&\cdots \qquad \cdots \qquad \cdots \qquad \cdots \nonumber \\
&\cdots \qquad \cdots \qquad \cdots \qquad \cdots \nonumber
\end{align}

The negative indexed conserved quantities,  namely eqs. \ref{negative-1},  \ref{negative-2} are  the members of negative 
hierarchy.
% and the first member  $H_{-1}$ gives the FLE.  

%From the above analysis on the phase dependent linear  soliton interference it is natural to expect that the same can be extended to multi soliton components.  

\section{Conclusion}
We have bilinearized  Fokas-Lenells equation(FLE) with a vanishing boundary condition. In the proposed bilinearization we have used  an  auxiliary function to convert the trilinear equations into a set of  bilinear equations. We have derived bright $1$-soliton, $2$- soliton solutions and presented  the scheme for obtaining $N$- soliton solution.  The additional parameter present in the soliton solution allows us to tune the position of  soliton.  We have  shown that with a suitable choice of parameters one soliton solution reduces to an algebraic soliton. We have also shown explicitly through asymptotic
analysis that the soliton interactions are elastic, that is the amplitude of each
soliton before and after interaction are same. The mark of interaction is left
behind only in the phase of each soliton.
Secondly we have proposed a generalised Lax pair for the FLE and obtained the conserved quantities by solving the Riccati equation to establish the integrability in the Liouville sense. We feel that the proposed scheme of bilinearization  considerably simplifies the procedure to obtain soliton solutions. 
We believe the present investigation would be useful to study the applications of FLE in nonlinear optics and other branches of physics.

\section*{Acknowledgement}
S Talukdar and R Dutta acknowledge DST, Govt. of India for Inspire fellowship,  grant nos.  DST/INSPIRE Fellowship/2020/IF200278 and DST/INSPIRE Fellowship/2020/IF200303.

\end{document}